\documentstyle[twocolumn,aps]{revtex}

\begin{document}

{\bf Comment on ``Gravity Waves, Chaos, and Spinning Compact
Binaries''}

\medskip

In {\cite{jl2000}}, Levin shows that the equations of motion for two
comparable mass objects in the second-post-Newtonian approximation to
general relativity predict chaotic orbits --- orbits whose future
evolution is extremely sensitive to initial conditions.  The phase
evolution of the gravitational waveforms emitted by these systems also
will be extremely sensitive to initial conditions.  This renders it
essentially impossible to detect such waves using the technique of
matched filtering.  The purpose of this Comment is to point out that
it has been understood for some time that alternative detection
methods will be needed in the relevant regime, so that matched
filtering is highly unlikely to be used for the detection of these
waves.

The post-Newtonian expansion of general relativity is a slowly
convergent, asymptotic expansion.  It is not very useful as the
members of the binary approach the innermost stable circular orbit.
The ``standard lore'' of the community studying sources of
gravitational waves (see, e.g., Ref.\ {\cite{btc1998}) is that
post-Newtonian approximations cease to be useful for constructing
accurate phase evolutions when the separation of the members of the
binary is $r \sim 15 M$.  Post-Newtonian phase evolution models can
describe the gravitational waveform when the separations are larger
than this; other methods adapted to the strong field (see, e.g., Ref.\
{\cite{bd1999}}) must be used for evolution inside this radius.  This
regime, beyond the realm of applicability of post-Newtonian
techniques, but prior to the point at which one would like to use
templates from numerical relativity, has been named the ``IBBH''
regime (see Ref.\ {\cite{btc1998} for further discussion).

It has been understood for some time that the waves emitted as compact
binary systems spiral through this regime are likely to be of great
importance for ground-based gravitational-wave observatories such as
LIGO, VIRGO, GEO600, and TAMA.  In particular, as binary black hole
(BBH) systems with total system masses of $10\,M_\odot\lesssim M
\lesssim 50\,M_\odot$ spiral through the IBBH regime, they emit waves
which lie roughly in the frequency band at which the first detectors
are maximally sensitive (see Ref.\ {\cite{fhI}}, particularly Fig.\
4).  Less massive systems (containing one or two neutron stars) emit
waves at higher frequencies, so that detectors are more sensitive to
earlier epochs of their coalescence, when post-Newtonian
approximations are probably reliable.

Because of the strong possibility that BBH coalescences will be common
events in the detectors' datastreams {\cite{pzm}}, and because it is
well-understood that theory is currently unable to reliably model the
phase evolution of their waves, much effort has been directed to build
data analysis techniques that do not rely on detailed phasing
information.  An example is the ``power-monitoring statistic''
introduced in Ref.\ {\cite{fhI}}, and discussed in greater detail in
Ref.\ {\cite{abcf}}.  Such techniques rely on the fact that the number
of cycles emitted as the system evolves through the very strong-field
regime is relatively small.  Matched filtering boosts the
signal-to-noise ratio by roughly the square root of the number of wave
cycles measured.  If the number of cycles is small, matched filtering
does not buy very much and --- at least for detection purposes ---
cruder techniques can do very well.

In Fig.\ 3 of {\cite{jl2000}}, Levin shows the fractal basin
boundaries of a binary system for many initial conditions, neglecting
dissipation due to gravitational wave emission.  This figure rather
clearly demonstrates the existence of chaos.  However, in this strong
field regime and with dissipation included, very few orbits remain
before the bodies merge, as is indicated by Fig.\ 5 of
{\cite{jl2000}}.  Methods such as the power-monitoring statistic would
probably do very well at detecting these waves.

Thus, I believe it is likely that the chaotic orbits of compact bodies
will not strongly impact the detection of gravitational waves by
ground-based observatories.  The number of cycles emitted during the
chaotic evolution Levin has demonstrated is probably small enough that
non-matched-filtering-based techniques are likely to work quite well,
particularly for early detectors whose noise characteristics render
them insensitive to all but the last few inspiral orbits (depending
rather strongly on the system's total mass).  Because post-Newtonian
templates are not accurate in this regime, it would have been very
surprising if matched filters were used to search for these waves even
in the absence of chaos.  It is worth noting, however, that chaotic
effects may dramatically impact our ability to extract astrophysical
information from the detected waves.  In addition, chaos may have a
big impact on detection of gravitational waves by LISA from the
inspiral of small ($\sim 1 - 10\,M_\odot$) bodies into massive ($\sim
10^5 - 10^7\,M_\odot$) black holes, where the number of cycles will be
very large ($\sim 10^5$).

\bigskip\noindent
Scott A.\ Hughes\\
130-33 Caltech\\
Pasadena, CA 91125\\
e-mail: hughes@tapir.caltech.edu

\end{document}